\documentclass[11pt,letterpaper,notoc]{JHEP}
\usepackage{epsfig}
\usepackage{cite}
\newcommand{\beq}{\begin{eqnarray}}% can be used as {equation} or {eqnarray}
\newcommand{\eeq}{\end{eqnarray}}
\newcommand{\nn}{\nonumber}
\def\ltap{\ \raise.3ex\hbox{$<$\kern-.75em\lower1ex\hbox{$\sim$}}\ }
\def\gtap{\ \raise.3ex\hbox{$>$\kern-.75em\lower1ex\hbox{$\sim$}}\ }

\def\CW{{\cal W}}

\def\be{\begin{equation}}
\def\ee{\end{equation}}
\newcommand{\bel}[1]{\be\label{#1}}
\newcommand{\beql}[1]{\beq\label{#1}}
\newcommand{\eref}[1]{(\ref{#1})}

\newcommand{\vev}[1]{ \left\langle {#1} \right\rangle }

\newcommand{\nomu}{{$\not\!\!\mu$SSM}}
\usepackage{latexsym}
\usepackage{epsfig}
\newcommand{\bc}{\begin{center}}
\newcommand{\ec}{\end{center}}
\newcommand{\y}{\nonumber \\}

\title{The Minimal Supersymmetric Model without a $\mu$ term}
\author{
Ann E. Nelson$^a$\footnote{anelson@phys.washington.edu}, Nuria Rius$^b$,  Veronica Sanz$^b$, Mithat Unsal$^a$ \hskip 0.2 cm\\
$^a$Department of Physics, Box 1560, University of Washington, 
                 Seattle, WA 98195-1560, USA\\
$^b$Departamento de F\'\i sica Te\'orica, IFIC, Universidad de Valencia-CSIC,
Edificio Institutos de Paterna, Apt. 22085, 46071 Valencia, Spain 
}
\preprint{UW/PT-02/05, IFIC/02-12, FTUV-02-0319}
\abstract{We  propose a supersymmetric extension of the standard model which is a realistic alternative to the MSSM, and which has several advantages.  
No ``$\mu$''  supersymmetric Higgs/Higgsino mass parameter is needed for 
sufficiently heavy charginos. An approximate $U(1)_R$ symmetry naturally 
guarantees that $\tan\beta$ is large, explaining the top/bottom quark mass 
hierarchy. This symmetry also suppresses supersymmetric contributions to 
anomalous magnetic moments, $b\rightarrow s \gamma$, and proton decay, and 
these processes place no lower bounds on superpartner masses, even at large  
$\tan\beta$. 
The soft supersymmetry breaking mass parameters can easily be obtained
from either gauge or Planck scale mediation, without the usual $\mu$ problem.
Unlike in the MSSM, there are significant {\it upper} bounds on the
masses of superpartners, including an upper bound of 114 GeV on the
mass of the lightest chargino. However the MSSM bound on the lightest
Higgs mass does not apply.}
\begin{document}
\section{Introduction}
\label{intro}
Supersymmetric theories with softly broken supersymmetry have no quadratically divergent contributions to the Higgs mass, and so supersymmetry is hailed as a solution to the gauge hierarchy problem.  In supersymmetric theories the $W$ and $Z$ masses are technically natural  provided the superpartners have mass in the vicinity of the weak scale.  

 However in the Minimal Supersymmetric Model (MSSM) of particle physics there is  a {\it supersymmetric} Higgs mass parameter,  ``$\mu$'',  which must be of order the   superpartner masses for successful phenomenology. It would seem to be a bizarre and unnatural coincidence that a supersymmetric mass should be of nearly the same size as the scale of supersymmetry breaking, unless both mass scales have a common origin. Most solutions to this ``$\mu$ problem'' focus on obtaining $\mu$ as a consequence of supersymmetry breaking, or obtaining   both supersymmetry breaking and the $\mu$ parameter from common inputs.  

In this paper we explore the consequences of a different approach to the $\mu$ problem, namely 
side-stepping this problem by building a viable model which does not have a $\mu$ parameter. We show that it is possible to obtain a spectrum of superpartner masses which is experimentally acceptable without $\mu$, provided the matter content of the MSSM is extended. In this model, which we call the ``$\mu$-less Supersymmetric Standard Model'' (\nomu), all mass arises directly from supersymmetry breaking or from electroweak symmetry breaking, and since the Higgs potential is determined exclusively from supersymmetry breaking terms, the electroweak scale is directly tied to the scale of superpartner masses. In the MSSM, where both $\mu$ and supersymmetry breaking terms contribute to the Higgs potential, there is the logical possibility of fine-tuning the $\mu$ parameter against supersymmetry breaking parameters to make the superpartners much heavier than the weak scale.  In the \nomu, such finetuning is not possible and the superpartner masses {\it must} be at the weak scale. 

In the next section we discuss the model and its spectrum. Remarkably, there is  an upper bound on the mass of the lightest chargino of 114 GeV,  not far beyond the kinematic reach of the  recently completed LEPII experiment and within reach of the TeVatron.

In section \ref{rsgmsb} we show how the supersymmetry breaking terms of the 
\nomu\ can arise from gauge mediation, when supersymmetry breaking in the 
messenger sector is mostly of the $D-$term type which respects an $U(1)_R$ 
symmetry. In section \ref{rssugra} we show that 
the approximate $U(1)_R$ symmetry is also  natural in the case of 
gravity mediated supersymmetry breaking, when the hidden 
supersymmetry breaking sector contains a gauge $U(1)$ with a 
a nonvanishing $D-$term and no gauge singlets.
In section \ref{T} we compute the leading one-loop contribution to the 
electroweak $T$ parameter.
In sections \ref{suppress} and \ref{pheno} we discuss some other desirable 
consequences of the approximate $U(1)_R$ symmetry: naturally large $\tan\beta$
without the usual 
enhancement of supersymmetric contributions to $g-2$, $b\rightarrow s \gamma$, and proton decay. 
In section \ref{unify} we discuss a messenger sector which renders the 
gauge mediated model 
compatible with gauge coupling unification, provided the messenger scale is 
less than $\sim 10^7$ GeV.

%----------------------------------------------------------------------------
\section{The \nomu\ model and its low energy spectrum}
\label{spectrum}
We  start with the principle that all mass terms arise directly either from electroweak symmetry breaking or from supersymmetry breaking. We therefore do not allow a supersymmetric $\mu$ term. The MSSM without a $\mu$ term  would have charginos lighter than the $W$ boson, which should have been found at LEP II, so we will have to extend the theory.
We  add the minimal matter content to the MSSM which will allow all charginos and visible neutralinos  to obtain mass beyond the current limits.   
What these current limits are is somewhat ambiguous since experimental limits are model dependent and have not been studied for this model. 
We will assume that charginos should be heavier than  104 GeV, the 
kinematic reach of LEP II. 

In the MSSM, the charginos arise from the charged spin 1/2 components of  the  Higgs superfields $H_2$ and $H_1$, and gauge  $W^\pm$ fields. With no $\mu$ term the mass matrix is

\bel{mssmchargino}
\begin{tabular}{|r|c  c|}
\hline
  & $-i \lambda^+ $ & $\Psi_{H_2}^+  $ \\
\hline
$-i \lambda^-   $ & $\tilde m_2 $ & $\sqrt{2} \,m_W \,s_{\beta} $ \\ 
 $\Psi_{H_1}^- $ &  $\sqrt{2} \,m_W \,c_{\beta} $  & $0$\\\hline
\end{tabular}
\ee
where $\tilde m_2$ is a supersymmetry breaking Majorana gaugino mass term and the off-diagonal entries break electroweak symmetry. For any value of $\tan\beta\equiv\vev{H_2/H_1}$ this mass matrix always has an eigenvalue less than the W mass and so is ruled out by the LEPII chargino mass bounds.

We will remedy this by adding matter which can mix with the MSSM charginos via supersymmetry breaking or electroweak symmetry breaking terms. The minimal such addition is a chiral superfield $T$ which transforms as a triplet under the $SU(2)$ gauge group, and is uncharged under the other gauge groups. We add a superpotential coupling
\bel{tcoupling}
\int d^2\theta\ \  h_T H_1 TH_2 \  .
\ee

Now the chargino mass matrix is 

\begin{equation}
\begin{tabular}{|r|c c c|}
\hline
  & $\Psi_T^+ $ & $-i \lambda^+  $ & $\Psi_{H_2}^+ $ \\
\hline
$\Psi_T^-  $ & $ 0$ & $\tilde M_2 $ & $ - 
h_T\, v_1$\\
$-i \lambda^-  $ & $ \tilde M_2$&$\tilde m_2$ & $\sqrt{2}\, m_W \,s_{\beta}$\\ 
$\Psi_{H_1}^-   $ & $h_T\, v_2$ &$\sqrt{2}\, m_W \,c_{\beta}  $&0\\ 
\hline
\end{tabular}
\label{chargino}
\end{equation}
where  $\tilde M_2$ is a  soft supersymmetry breaking  Dirac mass term. In the next sections we  will show how $\tilde M_2$ can be generated from gauge mediation or from hidden sector supersymmetry breaking. Note that  all the charginos can be made heavier than 104 GeV without a $\mu$ parameter. 

A potential problem with the superpotential coupling \eref{tcoupling}
is that the corresponding scalar trilinear can lead to a large
electroweak $T$ parameter. With a scalar trilinear $A_TH_2 T H_1$, the
Higgs vevs induce a tadpole for the $T$ scalar, which will then get a
vev.  An electroweak triplet vev is highly constrained by the electroweak $T$ parameter. The tadpole for the scalar may be suppressed in three ways. 
First,  the tadpole requires both the up and down type Higgs vevs and is suppressed at large $\tan\beta$. Second, the scalar trilinear is prohibited by $U(1)_R$ symmetry. Third, trilinear terms are rather suppressed even in conventional gauge mediated models, arising only at 2 loops.
All three suppressions are naturally present in a class of models
arising from the nearly $U(1)_R$ symmetric gauge mediation we discuss in 
section \ref{rsgmsb}, and the first two are present in the Planck
scale mediated models discussed in section \ref{rssugra}.  
Another potentially troubling scalar trilinear arises from the coupling of the triplet to the $SU(2)$ $D-$term,
$A_D T_a (H_2^\dagger t_a H_2 -H_1^\dagger t_a^* H_1)$, which is induced in both the models of supersymmetry breaking mediation we discuss.  Sufficient suppression of the resulting tadpole will be possible if the mass of the $T$ scalar is larger than a few TeV. In the gauge mediated models, a very large mass for the scalar, of order several TeV, is automatic. 

At this point the reader might worry that an approximate $U(1)_R$
symmetry will lead to a light pseudoscalar which is an approximate Goldstone boson. However we
will not  spontaneously break the symmetry (actually a linear
combination of the original $U(1)_R$ and electroweak hypercharge will
remain unbroken). This unbroken symmetry would require 
$\tan\beta\rightarrow\infty$, but we will explicitly break the symmetry by a small amount so that $H_2$ can get a small vev to give  the leptons and down-type quarks mass. Thus we can explain the top/bottom mass hierarchy via an approximate symmetry which gives naturally large $\tan\beta$ (as was also done in ref.\cite{Nelson:1993vc}).

We now turn to a discussion of the spectrum of the \nomu, from the bottom up. 
We start by giving the
charge assignments of some of  the  components of Higgs and electroweak
gauge fields under the unbroken $U(1)_R$:
\begin{equation}
\begin{tabular}{|r|c|}
\hline
$\Psi_{H_1}$&1\\
$\Psi_{H_2}$& -1\\
$\Psi_T^{\pm}$&-1\\
${H_1}$&2\\
${H_2}$&0\\
$\lambda^{\pm}$&1\\
\hline
\end{tabular}
\label{rcharge}
\end{equation}
It is also possible to assign $U(1)_R$ charges to quarks and leptons to allow  the usual MSSM superpotential coupling. 

With this $U(1)_R$  unbroken  the chargino mass matrix becomes
\begin{equation}
\begin{tabular}{|r|c c c|}
\hline
  & $\Psi_T^+ $ & $-i \lambda^+  $ & $\Psi_{H_2}^+ $ \\
\hline
$\Psi_T^-  $ & $ 0$ & $\tilde M_2 $ & $0$\\
$-i \lambda^-  $ & $ \tilde M_2$&0 & $\sqrt{2}\, m_W $\\ 
$\Psi_{H_1}^-   $ & $h_T\, v_2$ &$ 0 $&0\\ 
\hline
\end{tabular}
\label{rschargino}
\end{equation}
The eigenvalues of this matrix are:
\beq
m_{\chi^{\pm}_1} &=& \tilde M_2 \\
m_{\chi^{\pm}_{2,3}} &=&
\frac {1}{\sqrt{2}} \sqrt{\tilde M_2^2 + h_T^2 v^2 + 2 m_W^2 \pm 
\sqrt{(\tilde M_2^2 + h_T^2 v^2 + 2 m_W^2)^2 - 8 h_T^2 v^2 m_W^2}} \\
&\sim&
\sqrt{2} m_W \frac{h_T v}{\sqrt{h_T^2 v^2 + \tilde M_2^2}}\ , 
\sqrt{h_T^2 v^2 + \tilde M_2^2 + 
m_W^2 \frac{2 \tilde M_2^2}{h_T^2 v^2 + \tilde M_2^2}}
\eeq

This will get modified slightly by small $U(1)_R$ breaking effects, which will get us away from the limit $\tan\beta\rightarrow\infty$ and set $\tan\beta$ to a moderate value $\sim 60$.
In this limit there is one chargino with mass $\tilde M_2$ and another chargino whose mass decreases with $\tilde M_2$. To obtain masses for all charginos 
heavier than 104 GeV,  while assuming $h_T < 1.2$,  $\tilde M_2$ must be in the range 104-120 GeV.
Moreover, the requirement, that all charginos should be heavier than 104 GeV
 leads to a lower bound on the Yukawa coupling, $h_T \gtap 1$.
Note that $\sqrt2 m_W=114$ GeV is an upper bound on the mass of the lightest 
chargino. 
Thus in the region where all charginos are heavier than 104 GeV we have two 
charginos with mass between 104 and 120 GeV and one heavier one. 

We plot in Fig. 1 the lighter chargino masses as a function of 
$\tilde M_2$, for $h_T=1$, $\tan \beta =60$ and $\tilde m_2=$
5 GeV. 
In the range of $\tilde M_2$ where all charginos are heavier than 104 GeV
the mass of the heavier chargino is $\sim 270$ GeV.

\begin{figure} [t!] 
%\label{pchar}
\begin{center}
\epsfig{figure=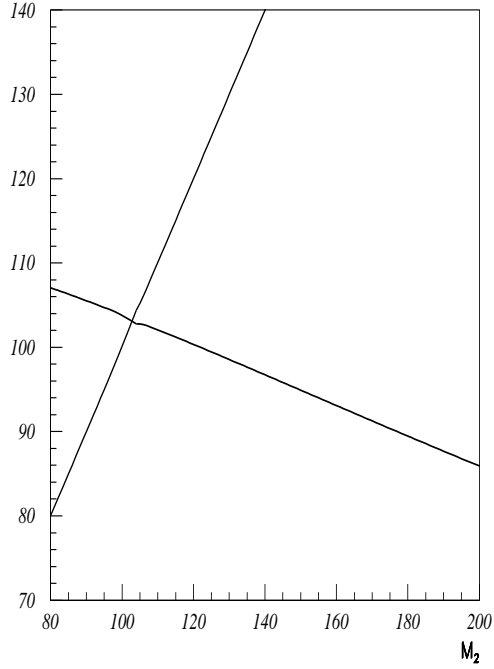,width=.5\textwidth,height=.5\textheight}
\end{center} 
\caption{Lighter chargino masses for $h_T=1$, $\tan \beta =60$ and 
$\tilde m_2=$5 GeV. }
\end{figure}

We now turn to the neutralino sector. In order to give the Bino a Dirac mass, 
analogous to the Wino mass term, we will add a singlet chiral superfield $S$ 
to the theory, and a coupling $h_S S H_1 H_2$ to the superpotential. This is 
reminiscent of the NMSSM, but in the present case the expectation value of 
the $S$  scalar will be much smaller than the electroweak scale. 
Alternatively, the singlet could be omitted, resulting in a nearly massless 
neutralino. 
Quantitative exploration of this more economical alternative has led us to the conclusion that it is difficult to simultaneously satisfy the constraints on the invisible width of the $Z$ and the $T$ parameter without the singlet, so we will describe
the theory with the singlet included.

The neutralino mass matrix is then:
\begin{equation}
\label{neutral}
\begin{tabular}{|r|c c c c c c|}
\hline
  & $\Psi_T^3$& $\Psi_S$&$-i \lambda'$ & $-i\lambda^3$ & $\Psi_{H_1}^1 $ &$\Psi_{H_2}^2$ \\
\hline
$\Psi_T^3$ & $0$ &0& $0$ & $ \tilde{M_2} $ & $h_T\, v_2/\sqrt{2}$ & 
$h_T \,v_1/\sqrt{2}$\\
$\Psi_S$&0&$0$&$\tilde{M_1}$&0&$h_S\,v_2/\sqrt{2}$&$h_S\,v_1/\sqrt{2}$\\
$-i \lambda'$ & $0$ &$\tilde{M_1}$& $\tilde m_1$ & $0$ & $-m_Z \, s_W \, c_{\beta}$ & $m_Z \, s_W \, s_{\beta} $\\
$-i\lambda^3$  &$ \tilde{M_2} $ &0& $0$ & $\tilde m_2$ & $m_Z \, c_W \, c_{\beta}$ & $-m_Z \, c_W \, s_{\beta} $\\
$\Psi_{H_1}^1$ & $h_T \, v_2/\sqrt{2}$ &$h_S\,v_2/\sqrt{2}$& $-m_Z \, s_W \, c_{\beta}$ & $m_Z \, c_W \, c_{\beta}$ & $0$ &
$0$\\ 
$\Psi_{H_2}^2$ & $h_T \, v_1/\sqrt{2}$ &$h_S\,v_1/\sqrt{2}$& $m_Z \, s_W \, s_{\beta}$ & $-m_Z \, c_W \, s_{\beta}$ & $0$ &
$0$\\ 
\hline
\end{tabular}
\end{equation}

In the large $\tan\beta$, $U(1)_R$ symmetric limit the masses become 
approximately Dirac, with a mass matrix of the form

\begin{equation}
\begin{tabular}{|r|c c c  |}
\hline
  & $\Psi_T^3$& $\Psi_S$ &$\Psi_{H_2}^2$ \\
\hline
$-i \lambda'$ & $0$  & $\tilde M_1$ & $m_Z \, s_W  $\\
$-i\lambda^3$  &$\tilde M_2$ & 0 & $-m_Z \, c_W $\\
$\Psi_{H_1}^1$ & $h_T \, v_2/\sqrt{2}$ & $h_S\,v_2/\sqrt{2} $& 0  \\ 
\hline
\end{tabular}
\label{rsneutralino}
\end{equation}

Note that in this limit there is always a nearly Dirac neutralino with mass 
lighter than the $Z$.
In Fig. 2 we show the  neutralino masses as a function of the
soft mass term $\tilde M_1$, for $\tilde M_2 = 104$ GeV, $h_T = 1$ and 
$h_S = 0.1$. In principle the Yukawa coupling $h_S$ is a free parameter, 
but large values are disfavored by electroweak precision measurements 
(see section \ref{T}).
We have taken all Majorana gaugino masses equal to 5 GeV and 
$\tan \beta =60$. 

Notice that in certain regions of parameter space the lightest 
quasi-Dirac neutralino is lighter than $m_Z/2$, thus the decay of
the $Z$ to these neutralinos is kinematically allowed. 
The resulting increase in the width of the $Z$ is typically very small, 
due to the fact that the lightest neutralino has only a small 
higgsino component.

\begin{figure} [t!] 
%\label{pneu}
\begin{center}
\epsfig{figure=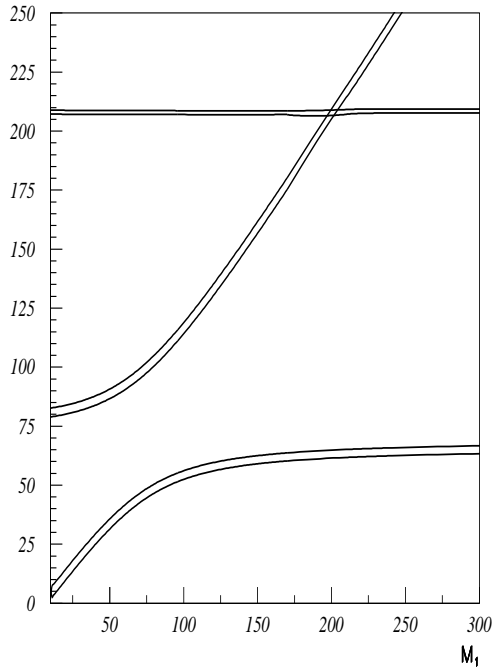,width=.5\textwidth,height=.5\textheight}
\end{center} 
\caption{Neutralino masses as a function of $\tilde M_1$, for 
$\tilde M_2 = 104$ GeV, $h_T = 1$, $h_S = 0.1$, $\tan \beta=60$ and 
$\tilde m_1=\tilde m_2=$5 GeV.}
\end{figure}

In order to allow a gluino mass without breaking the R-symmetry we
also introduce a chiral superfield $O$, which is a color octet, and a
supersymmetry breaking  Dirac mass term
\be
\tilde M_3 \psi_O \lambda_8
\ee
 where $\lambda_8 $ is the gluino field.

The scalar superpartners receive soft supersymmetry breaking masses as
usual. Trilinear scalar couplings  are suppressed by $U(1)_R$ and
quite small.
A very small scalar $\mu B$ term of order a few GeV$^2$
\be \mu B H_1 H_2\ee will be needed in order to induce a small vev for
$H_1$. It is natural for this term to be small as it breaks the
approximate $U(1)_R$ symmetry. Because the symmetry is explicitly broken rather than spontaneously broken, there is no light pseudoscalar. Similarly, in the MSSM, small $\mu B$ does not lead to a light pseudoscalar when $\tan\beta$ is  large.

Speculations on the origin of these supersymmetry breaking terms are the 
subject of the next two sections.

%----------------------------------------------------------------------------

\section{$U(1)_R$ Symmetric Gauge Mediation}
\label{rsgmsb}
In this section we assume that supersymmetry breaking is transmitted to
the \nomu \ by Gauge Mediated Supersymmetry Breaking (GMSB).
As usual, we assume a messenger sector of heavy supermultiplets in a vector-like
representation of the standard gauge group. 
In conventional gauge mediation, the messengers learn about
supersymmetry breaking from coupling to a gauge singlet with an $F-$term.
This transmits both supersymmetry breaking and $U(1)_R$ symmetry
breaking to the MSSM. Since we want an approximately $U(1)_R$
symmetric \nomu, we will assume the messenger sector does not contain any singlet.  Instead supersymmetry breaking in the
messenger sector is primarily mediated by some new
gauge group also carried by the messengers. Such mediation will primarily induce nonholomorphic scalar supersymmetry breaking masses in the messenger sector. Note that it is simple to construct theories with new
  gauge interactions carried by the messenger fields  which produce
  such soft  masses in the messenger sector 
\cite{Randall:1997zi,Csaki:1998if}. In fact, such theories are even simpler and more natural than most conventional gauge mediated models, which generally require some complicated model building in order to induce the required singlet $F-$term and messenger mass scale.

For simplicity,  in this section we  assume the usual 
messenger matter content of chiral superfields $L,\bar L, D,\bar D$
where $L,\bar L$ transform under $SU(2)\otimes U(1)$ in conjugate
representations  and $D, \bar D$
carry color.  In section \ref{unify} we will give
examples of representations which allow for successful coupling
constant unification.

In order to obtain Dirac gaugino masses, $S$, $T$ and $O$ must couple to
the messengers. The messenger superpotential   
is
\bel{mess}
\lambda_S S \bar L L+\lambda'_S S \bar D D +
\lambda_T T \bar L L+ \lambda_O O\bar D D + M_L \bar L L + M_D \bar D D
\ .
\ee
The supersymmetric mass parameters $M_L$ and $M_D$ can be much heavier
than the weak scale, and we will not discuss their origin here.

The mass matrix for, {\it e.g.} the $L, \bar L$ scalar fields will
have the following  form 
\bel{messass}
\pmatrix{
M_L^2 + \tilde m_L^2 & 0\cr  
0 & M_L^2 +\tilde m_{\bar L}^2  
}
\ee
where $\tilde m_L^2,\tilde m_{\bar L}^2$ are soft supersymmetry
breaking masses. 
 However, with no messenger singlet, to leading order the messenger sector will accidentally have unbroken $U(1)_R$ symmetry, and no
Majorana gaugino masses will be produced.  In the models of refs. \cite{Randall:1997zi,Csaki:1998if}, the lack of one loop Majorana gaugino masses was a phenomenological problem. In the \nomu, however, at one loop, 
the gauginos couple to the fermionic components of $T,O$ and $S$ and get a Dirac
supersymmetry breaking mass.

 Note also that provided the $D-$type masses are generated by new gauge interactions whose generators are orthogonal to electroweak hypercharge,  {\it i.e.}
\be {\rm Tr}\ T_Y T_{\rm new}=0\ ,\ee 
the  disaster of generating a $D$-term for hypercharge at one 
loop is avoided.

There are two diagrams contributing to Dirac gaugino masses, which cancel in
the limit that  $\tilde M_{L,D}^2= \tilde M_{\bar L,\bar D}^2$.

 Defining the mass-squared ratio of scalar particles to fermion
as 

\beq 
y_{L,D}&=& \frac{M_{L,D}^2+\tilde m_{L,D}^2}{M_{L,D}^2} \nn\\
\bar y_{L,D}&=& \frac{M_{L,D}^2+\tilde m_{\bar L,\bar D}^2}{M_{ L,D}^2}\\
\eeq
we find Dirac masses $\tilde M_{2,3}$ of
\be
\tilde M_{2,3}=S_{L,D}\,  M_{L,D}\, \frac{g_{2,3} \lambda_{T,O}}{2\pi^2}\left[\frac{y_{L,D}
  \log(y_{L,D})}{1- y_{L,D}} -\frac{\bar y_{L,D}
\log(\bar y_{L,D})}{1- \bar y_{L,D}}\right]   
\ee
 where $S_{L,D}$ are the Dynkin indices of the
$L,D$ representations respectively.  Similarly, $\tilde M_1$ will receive contributions from both $L$ and $D$.

In the limit  that the supersymmetry breaking terms are much smaller
than $M_L$, the result is
\be
\tilde M_{2,3}=S_{L,D}   \frac{g_{2,3} \lambda_{T,O}}{4\pi^2} { \tilde m_{\bar
    L, \bar D}^2-\tilde m_{L,D}^2\over M_{L,D}}\ .
\ee

If the mass squared differences  are regarded as arising from a $D$ component of a $U(1)$ gauge field $\CW'$, then these graphs may be regarded as generating a supersymmetric operator
\be
\label{supersoft}
\int d^2\theta {\xi_1\over M} \CW^\alpha_1\CW'_\alpha S \ + \ {\xi_2\over M} \CW^\alpha_2\CW'_\alpha T \ + \ 
{\xi_3\over M}\CW^\alpha_3 \CW'_\alpha O\  ,
\ee
where $M$ is the messenger mass scale, $\CW_{1,2,3}$ are the standard model $U(1)$, $SU(2)$ and $SU(3)$ gauge
field strengths and $\xi_{1,2,3}$ are dimensionless numbers. Note that supersymmetry breaking scalar trilinear couplings of the $T,O,S$ scalars to the $D$ components of the MSSM gauge fields, which are contained in the 
operator (\ref{supersoft}), are also generated by gauge mediation, as are the operators
\be
\label{lemontwist}
\int d^2\theta {{\CW'}^\alpha \CW'_\alpha\over M^2} ( \xi_1' S^2+\xi_2' T^2+\xi_3' O^2)\ .
\ee
The operators (\ref{lemontwist}) will give  masses squared  of opposite sign to the scalar and pseudoscalar components of $S,T,O$.
Because of the large positive  gauge mediated mass for the spinless components of $S,T,O$, an additional negative contribution to one of the masses squared is not troublesome.

The masses of scalar \nomu\ particles may be found as a special case of
the general expressions computed in \cite{Poppitz:1997xw,Arkani-Hamed:1998kj}.
Note that obtaining positive squark and slepton  masses will require negative
supertrace in the messenger sector, i.e \footnote{An alternative possibility for squark and slepton masses is to generate them from finite loops involving the Dirac gaugino mass \cite{fox}. However this would require ultra heavy gauginos of order several TeV, which is incompatible with the \nomu.}
\beq
\tilde m_{\bar L}^2+\tilde m_{L}^2&<&0\\
\tilde m_{\bar D}^2+\tilde m_{D}^2&<&0.
\eeq

With a negative supertrace of messenger sector masses squared, the scalar components of $T,S,$ and $O$ will receive a large positive mass squared at one loop and will therefore be significantly heavier than the other superpartners. This mass is of order a loop factor times the soft masses in the messenger sector, and is not suppressed by the messenger mass scale. The $T$ and $O$ scalar masses  should not be much larger than
$10^4$ GeV, or they  will give excessive two loop  contributions to squark and slepton masses.  The supersymmetry breaking terms in the messenger sector should therefore not be larger than of order $M_S\sim 10^5$ GeV.  Since squark and slepton masses will be of order $(\alpha/\pi)(M_S^2/M)$, the messenger mass scale $M$ should  be below $10^6$ GeV.

Gauge mediated models have a generic $\mu$ problem.  It is quite
difficult  in gauge mediation to induce a $\mu$ parameter which is naturally related to supersymmetry breaking,  without inducing an excessively
large
$B\mu$ parameter \cite{Dine:1995vc}. The \nomu\ avoids the gauge mediated $\mu$
problem. A  $\mu B$ parameter can be induced which is proportional to a small coupling,  
 and it is not a problem that the resulting $\mu$ parameter will be much smaller than the weak scale.

It is  a simple matter to
induce the very small  $B\mu$ term needed for $\tan\beta < \infty$. 
For instance the messenger
sector could contain a very heavy gauge singlet field $S'$ with a coupling
$S'H_1H_2$. If multi loop effects induce a small vev and  $F$-term for $S'$ 
from the
supersymmetry breaking in the messenger sector the requisite $B\mu$
term can be induced. The $\mu$ term from $\vev{S'}$ will be quite small, 
and of no phenomenological importance.

%----------------------------------------------------------------------------

\section{\nomu\  with  Hidden Supersymmetry Breaking}

\label{rssugra}
The \nomu\ model with the requisite approximate accidental $U(1)_R$
symmetry may also arise naturally in  hidden sector models where the dominant
mediation mechanism is from  Planck scale physics. In this section
we will show that the approximate $U(1)_R$ is automatic
when the hidden  supersymmetry breaking sector does not contain any 
gauge singlets, but does have a gauged $U(1)$ with a nonvanishing
$D-$term. It is quite simple to build such models with dynamical supersymmetry breaking.
Gravity mediated $U(1)_R$ symmetric
supersymmetry breaking has  been considered before  
\cite{Hall:1991hq,Randall:1992cq,Dine:1992yw,Feng:1996dn}. Refs. \cite{Hall:1991hq, Randall:1992cq} introduced the $O$ but not the
 $T$ and $S$ fields, and always predicted
a chargino lighter than the $W$, hence are  now
now ruled out. Ref. \cite{Dine:1992yw} introduced also a pair of $T$ fields, and mentioned a possible supergravity origin for  Dirac mass term in theories with hidden dynamical supersymmetry breaking.

Following ref. \cite{Dine:1992yw}, we assume that a hidden supersymmetry breaking sector contains some
gauge interactions, including a $U(1)$ with a gauge field strength
$\CW_\alpha$ which has nonvanishing $D-$term, and chiral superfield(s) $X$ which are  charged under the hidden sector gauge symmetry,  which 
obtain $F$-terms. The `4-1' \cite{Dine:1996ag,Poppitz:1996fh} theory of dynamical
supersymmetry breaking is a simple example of such a model.

We allow the most general gauge invariant nonrenormalizable interactions between the
hidden and the \nomu\ visible sectors (consistent with enough approximate flavor
symmetry to adequately suppress flavor changing neutral currents). Since there is no $X$ gauge singlet, there are no gauge invariant terms linear in $X$, hence any Majorana gaugino masses or  trilinear scalar couplings will be  suppressed. The
Dirac gaugino masses arise from

\bel{dgauginomass}
\int d^2\theta {\xi_1\over M_P}\CW^\alpha_1\CW'_\alpha S \ + {\xi_2\over M_P}\CW^\alpha_2\CW'_\alpha T \ + \ 
{\xi_3\over M_P}\CW^\alpha_3 \CW'_\alpha O \ .
\ee
The D component of $\CW'$, which we call $D'$, will give 
\beql{diracorigin}
\tilde M_1&= \xi_1 {D'\over M_P}\\
\tilde M_2&= \xi_2 {D'\over M_P}\\
\tilde M_3&= \xi_3 {D'\over M_P}
\eeq
Scalar masses  squared can arise in the usual way from the
operators
\bel{scalarmass}
\int d^4\theta {\xi_0\over M_P^2} X^\dagger X Q^\dagger Q
\ee
where $Q$ is an \nomu\ chiral superfield.
For $F_X\sim D'$ these should be the same approximate size as the
gaugino masses.

There will also be an  anomaly mediated \cite{Randall:1998uk, Giudice:1998xp} contribution to
Majorana gaugino masses and scalar trilinears. These effects are down by a loop factor but
 will  lead to a small amount of  $U(1)_R$ symmetry breaking.

The $B\mu$ term can arise from the coupling
\bel{bmu}
\int d^4\theta {\xi_b\over M_P^2} X^\dagger X H_1 H_2\ .
\ee

There is no necessary reason why this term should be smaller than the
electroweak scale, but as it breaks the R-symmetry it is quite natural in the sense of 'tHooft that it
should be small. Note that a $\mu$ and a $B\mu$ term with $B=m_{3/2}$  will also arise if $H_1 H_2$
appears in the Kahler potential \cite{Randall:1998uk} but
again   an approximate  R symmetry makes  it  natural for this term to be small.

Kinetic mixing between hypercharge and the hidden $U(1)$ must be suppressed, as it would lead to a large hypercharge $D-$term. Also potentially dangerous
are terms linear in the singlet $S$ such as
\be\int d^2\theta\ \CW'^\alpha\CW'_\alpha S +\ldots\ee
which could give the singlet scalar a large tadpole.
The necessary suppressions can be guaranteed by symmetries and nonrenormalization theorems, provided hypercharge is unified into a nonabelian group, and provided $S$, along with $T$ and $O$, is part of an adjoint of the unified group.

\section{SUSY contributions to precision electroweak parameters}
\subsection{$T$ parameter}
\label{T}

The approximate $U(1)_R$ symmetry of the
\nomu\ model and/or a heavy mass for the triplet scalar provides sufficient suppression of the tree level 
contribution to the electroweak $T$ parameter. However the 
superpotential couplings $h_T T H_1 H_2$ and $h_S S H_1 H_2$
break custodial $SU(2)$ symmetry and thus can lead to potentially large 
one-loop effects in the $T$ parameter. One should keep in mind that the
oblique approximation is not appropriate for light superpartners, 
and the complete supersymmetric 
one-loop corrections in this model should be considered. 
Such a computation is  beyond the scope of the present work, so we
shall interpret our results for the $T$ parameter as an order of
magnitude estimate of the radiative corrections expected in the 
\nomu\ model. As we will see, the effects can be sizable. 

We work in the large $\tan \beta$ limit. The 
main contributions to the $T$ parameter come from chargino
and neutralino loops in the $Z$ and $W$ self-energies.
The complete expression is rather cumbersome and can be evaluated only
numerically, however we have found that the approximation 
of keeping just the entries proportional to $h_T, h_S$
in the chargino and neutralino mass matrices is quite good and 
leads to simple analytic results.
In this limit we obtain 
\beq
T&= &\left(\frac{h_T^2 v^2}{32 \pi m_Z^2 s_W^2 c^2_W (1-x)}\right) \times\nn\\
&&\left[-5 + 6 x - x^2
-2 (x^2+2 x+5) \log\left(\frac{h_T^2 (1+x) v^2}{2 \mu^2}\right)
+ 4(x+3) \log\left(\frac{h_T^2 v^2}{\mu^2}\right)\right] 
\label{tpar}
\eeq
where $x=h_S^2/h_T^2$ 
and the renormalization scale $\mu$ should be taken to be $m_Z$.

Eq. (\ref{tpar}) shows that the leading contribution to the $T$ 
parameter grows as $h_T^2 \log (h_T^2 v^2/\mu^2)$ and it is 
therefore very sensitive to the exact value of the coupling $h_T$.
Recall that there is a lower limit on this coupling from chargino
masses. Although the singlet coupling $h_S$ also contributes
to the $T$ parameter, its contribution is negligible provided 
$h_S \ltap 0.1$ and we will ignore it in the following.

There is also a $T$ parameter contribution from the scalar sector, due
to the mass splitting between the different $SU(2)$ components 
of the Higgs doublet $H_1$ and the triplet $T$.
This contribution is not enhanced by the $\log \mu^2$ term though,
and can be made very small by the soft supersymmetry breaking scalar
masses.

From a global fit of the electroweak precision data one obtains 
$T=-0.02 \pm 0.13 (+0.09)$, where the central value assumes 
$M_H= 115$ GeV and the parentheses shows the change for $M_H =300$
GeV \cite{Groom:2000in}. 
This bound can be relaxed for larger $M_H$, leading to
$T \ltap 0.6$ at 95\% CL \cite{Chivukula:2000fe}.

If we impose the kinematic limit from LEP II that charginos should be
heavier than 104 GeV, $h_T \sim 1$ and the contribution to the $T$ 
parameter is  huge, $\sim$ 2.7.
However if the actual bound on chargino masses in this model were 
somewhat lower, say 90 GeV, we would obtain $h_T \gtap 0.6$
which leads to $T \sim 0.6$. 
Therefore, given the large sensitivity of the $T$ parameter to the 
value of $h_T$, a careful calculation of the chargino mass bounds  
is crucial to determine the viability of the \nomu\ model. 

Moreover, the determination of the $T$ parameter comes mainly from
the $Z$ width. As we have mentioned in section \ref{spectrum},
in certain regions of parameter space the $Z$ can decay to the lightest 
nearly Dirac neutralinos, and the above bounds on the $T$ parameter 
will not directly apply.

\subsection{Muon anomalous magnetic moment}
%$g_{\mu}-2$}
\label{suppress}

In the \nomu\ model, the approximate $U(1)_R$ symmetry suppresses the 
supersymmetric contributions to anomalous magnetic moments, electric dipole moments,
$b\rightarrow s \gamma$, and proton decay. As an example, we have 
explicitly computed the contribution to the anomalous magnetic 
moment of the muon, $a_\mu$. 
The measurement of $a_\mu$ by the $g-2$ collaboration \cite{Brown:2001mg} is consistent with but 
slightly higher than the Standard Model prediction, 
$a_\mu^{exp} - a_\mu^{SM} = (21\pm 18) 10^{-10}$, if one uses 
$a_\mu^{had}=(697 \pm 10) 10^{-10}$ for the hadronic polarization
contribution \cite{Davier:1998si,Jegerlehner:2001ca} and 
$(8\pm 3) 10^{-10}$ for the hadronic contribution to light by light scattering.  The hadronic contribution to light by light scattering has been calculated in \cite{Hayakawa:2001bb,Knecht:2001qg,Knecht:2001qf,Marciano:2001qq,Bartos:2001pg,Bijnens:2001cq,Blokland:2001pb}, with satisfactory agreement between the calculations. However there is intrinsic theoretical uncertainty and hadronic model dependence in these calculations, due to nonperturbative hadronic physics, as emphasized in \cite{Melnikov:2001uw}. A model independent  effective chiral Lagrangian treatment includes a low energy constant which is not constrained by other data \cite{Ramsey-Musolf:2002cy}. The size of this term may be estimated from the model calculations. Assuming the size  of the term is of the order indicated by modeling the hadronic physics gives an intrinsic theoretical uncertainty of order $3 \times 10^{-10}$ in the hadronic contribution to the light by light.
However, as emphasized in \cite{Ramsey-Musolf:2002cy},  errors in estimating the parameters which account for nonperturbative effects are not gaussian and a factor of three or more deviation from the estimate of the low energy constant would not be unusual.

The supersymmetric contributions to $a_\mu$ 
\cite{Moroi:1996yh,Martin:2001st} include  
loops with a chargino and a muon sneutrino and loops with a 
neutralino and a smuon. 
Besides the chargino and neutralino mass matrices given in section 
\ref{spectrum}, we also need the 
smuon mass matrix. Note that the trilinear A is expected to be small in this model. We will take it to be real, as would occur it it is generated by anomaly or gauge mediation.
\begin{equation}
M_{\tilde \mu}^2 = 
\left(
\begin{array}{cc}
m_{L}^2+(s_W^2-\frac{1}{2}) \, m_Z^2 \,  c_{2\beta} & A \, v_1\\
A \, v_1 & m_{R}^2-s_W^2 \, m_Z^2 \,  c_{2\beta}
\end{array}\right),
\label{smuons}
\end{equation}
and the sneutrino mass 
\begin{equation}
m_{\tilde{\nu}}^2=m_{L}^2+\frac{1}{2}\,  m_Z^2 \,  c_{2\beta} \ ,
\end{equation}
where we have used the notation $c_{2\beta} \equiv \cos 2\beta$.

Performing the summation over all chargino, 
neutralino and smuon mass eigenstates, the result for $a_\mu$ reads
\beq 
\delta a_\mu^{\chi^0} &=&  \frac{m_\mu}{16 \pi^2} \sum_{i,m}
\left\{- \frac{m_\mu}{12 m_{\tilde \mu_m}^2}
(|n_{im}^L|^2 + |n_{im}^R|^2) F_1^N(x_{im}) + 
\frac{m_{\chi_i^0}}{3 m_{\tilde \mu_m}^2} {\rm Re}[n_{im}^L n_{im}^R]
F_2^N(x_{im}) \right\}
\label{deltan} 
\\
\delta a_\mu^{\chi^{\pm}} &=&  \frac{m_\mu}{16 \pi^2} \sum_{k}
\left\{\frac{m_\mu}{12 m_{\tilde \nu_\mu}^2}
(|c_k^L|^2 + |c_k^R|^2) F_1^C(x_k) + 
\frac{2 m_{\chi_k^{\pm}}}{3 m_{\tilde \nu_\mu}^2} {\rm Re}[c_k^L c_k^R]
F_2^C(x_{k}) \right\}
\label{deltac} 
\eeq
where $i=1,\ldots,6$, $k=1,2,3$ and $m=1,2$ are neutralino, chargino and 
smuon mass eigenstate labels respectively, 
$x_{im}={m_{\chi_i^0}^2}/{m_{\tilde \mu_m}^2}$ and
$x_k={m_{\chi_k^{\pm}}^2}/{m_{\tilde \nu_\mu}^2}$.
The interaction vertices $n_{im}^{L,R}, c_{k}^{L,R}$ and the loop 
functions $F_i^{N,C}(x)$ can be found in the appendix.

In the $U(1)_R$ symmetric limit, the contributions to $a_\mu$ proportional 
to the neutralino and chargino masses exactly vanish, and there is only
a tiny effect proportional to $m_\mu$. However, 
once we take into account the small $U(1)_R$ symmetry breaking effects,
the leading contribution comes from the terms with the neutralino and 
chargino masses, much as in the MSSM.
There are two kinds of corrections, approximately of the 
same order: terms proportional to the gaugino Majorana masses, 
$\sim $ few GeV and terms proportional to 
$v_1 (\sim $ 4 GeV for $\tan \beta = 60$). 

\begin{figure} [ht!] 
%\label{gminustwo}
\begin{center}
\epsfig{figure=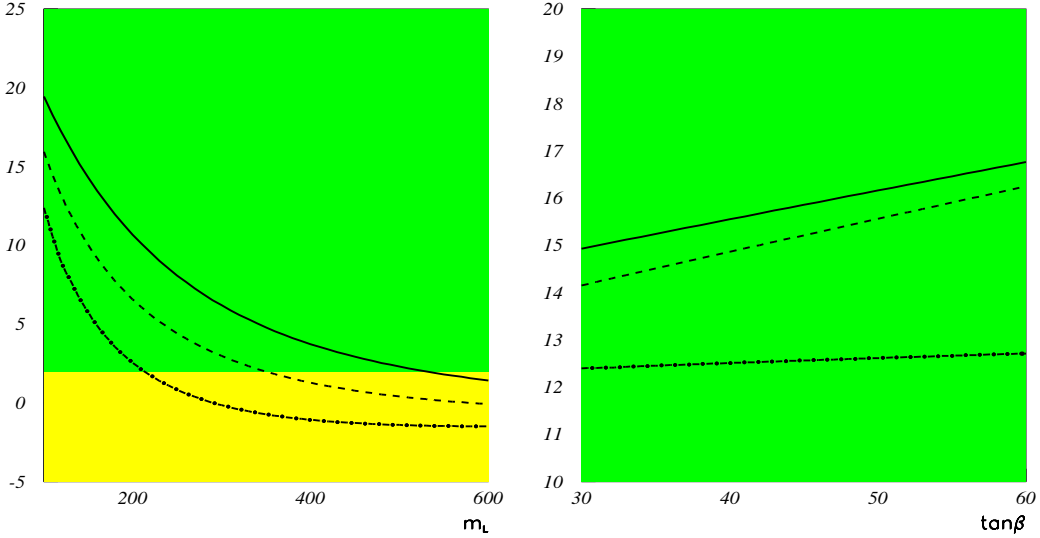,width=\textwidth,height=.4\textheight}
\end{center} 
\caption{Maximum value of $\delta a_{\mu} \times 10^{10}$ as a function of 
$m_L$ and $\tan \beta$,  
for
$ \tilde{m}_1 =\tilde{m}_2$ = 0 (dashed-dotted), 
5 GeV (dashed) and 10 GeV (solid).
We have taken $A=0$, $m_R =$ 100 GeV, $\tilde{M}_1=100$ GeV,  
$\tilde{M}_2=110$ GeV, $h_T = 0.8,\ h_S=0.1$, 
 $\tan\beta=60$ (left) and $m_L =$ 100 GeV (right).
The shadowed areas correspond to 1$\sigma$ (dark-green) and 2$\sigma$ 
(light-yellow) allowed regions from the $g-2$ collaboration result.}
\end{figure}

In Fig. 3 (left) we show the maximum possible value of 
$\delta a_\mu$ in the \nomu\ model as a function of the soft 
supersymmetry breaking mass term $m_L$,  
for several values of the gaugino Majorana masses. We have taken 
all of them equal, but the results are not very sensitive 
to this particular choice. 
We have fixed $\tilde M_1$ = 100 GeV, $\tilde M_2$ = 110 GeV, $A=0$,  
$m_R =$ 100 GeV and $\tan \beta =60$. 
These values of the soft supersymmetry breaking parameters are consistent 
with the constraint that all charged superpartners are heavier than 
100 GeV. 
In Fig. 3 (right) we have plotted the maximum possible value of 
$\delta a_\mu$ as a function of $\tan \beta$, 
for $m_L =$ 100 GeV and the same values of the remaining parameters. 

As we discussed in sec. \ref{spectrum}, the allowed range of $\tilde M_2$, 
$h_T$ and $h_S$ for successful phenomenology is quite constrained, so 
there is not significant dependence of $\delta a_\mu$ on these parameters. 
We also observe that the neutralino contribution 
 is approximately independent of $\tilde M_1$.

Regarding the dependence of $\delta a_\mu$ on the soft masses 
$m_L, m_R$, we find that for $m_L \approx m_R$
the contribution from chargino loops is typically larger
than the neutralino one by about one order of magnitude.
If $m_R \gg m_L$, chargino-sneutrino loops are still dominant, since 
the sneutrino mass depends only on $m_L$, while if $m_L \gg m_R$, 
the chargino-sneutrino loops rapidly decrease, and 
both contributions become comparable
\footnote{Notice that in the \nomu\ model the
trilinear terms are very small, and moreover they are multiplied 
by $v_1$ in the smuon mass matrix (\ref{smuons}), so the mixing in the
smuon sector is negligible.}. 
For instance, if $m_R =$ 100 GeV and $m_L \ge$ 500 GeV, then 
$\delta a_\mu^{\chi^0} \approx \delta a_\mu^{\chi^{\pm}}$ and 
as we see in Fig. 3, the total 
$\delta a_\mu$ is about one order of magnitude smaller than 
for $m_L=m_R =$ 100 GeV.

We have performed an analytic approximation, in order to better 
understand the numerical results. Since the chargino contribution is
typically dominant we focus on this one. 
We have diagonalized the chargino mass matrix in the large $\tan \beta$
limit. Then we consider the $U(1)_R$ symmetry breaking entries as
perturbations, and diagonalize the complete matrix to first order in
the small parameters $\tilde m_2, v_1$. 
We obtain  
\be
\delta a_{\mu}^{\chi^{\pm}}= - \frac{\sqrt{2} g_2}{24 \pi^2}
\frac{m_{\mu}^2}{m_{\tilde \nu_\mu}^2} 
\{ A \tan \beta  +  B \} \ , 
\label{aapprox}
\ee
where in the limit $m_W \ll h_T v, \tilde M_2$ the coefficients 
$A,B$ are given by 
\be
A =  \tilde{m}_2 h_T \frac{\tilde M_2}{m_{\chi^{\pm}_3}^2} [
G(m_{\chi^{\pm}_1},m_{\chi^{\pm}_3}) - G(m_{\chi^{\pm}_1},m_{\chi^{\pm}_2})] 
\ee
and 
\beq
B &=& \frac{g_2}{\sqrt{2}} \left(
\frac{\tilde M_2^2}{m_{\chi^{\pm}_3}^2}
G(m_{\chi^{\pm}_1},m_{\chi^{\pm}_2}) +
\frac{h_T^2 v_2^2}{m_{\chi^{\pm}_3}^2} G(m_{\chi^{\pm}_1},{m_{\chi^{\pm}_3}})
\right)
\nonumber \\  
&+&h_T\frac{\tilde M_2}{m_{\chi^{\pm}_3}}
F(m_{\chi^{\pm}_1},m_{\chi^{\pm}_2}) 
- \frac{g_2 h_T^2}{\sqrt{2}} \frac{\tilde M_2 v_2^2}{m_{\chi^{\pm}_3}^3}
F(m_{\chi^{\pm}_1},m_{\chi^{\pm}_3}) 
\eeq
where we have defined
\beq
F(m_{\chi^{\pm}_1},m_{\chi^{\pm}_i})&=&
\frac{m_{\chi^{\pm}_1} m_{\chi^{\pm}_i}}
{m_{\chi^{\pm}_i}^2 -m_{\chi^{\pm}_1}^2} 
\left[ F_2^C(x_i) - F_2^C(x_1) \right] \ ,
\\
G(m_{\chi^{\pm}_1},m_{\chi^{\pm}_i})&=&
\frac{m_{\chi^{\pm}_i}^2 F_2^C(x_i) -  m_{\chi^{\pm}_1}^2 F_2^C(x_1)}
{m_{\chi^{\pm}_i}^2 -m_{\chi^{\pm}_1}^2}\ .
\eeq

From the approximate expression (\ref{aapprox}) we see that 
although the contribution to $a_\mu$ in the \nomu\ model 
is also enhanced for large $\tan\beta$, due to the approximate $U(1)_R$ it is suppressed by the small
gaugino Majorana masses and therefore much smaller than in the
MSSM.

\section{Proton Decay}
\label{pheno}
One advantage of a $U(1)_R$ symmetry is that a large  ratio of Higgs vevs $H_2/H_1$ becomes  natural, explaining the ratio of top to bottom quark masses in terms of an approximate symmetry. Thus a large number of supersymmetric processes which are normally dangerously large in the large $\tan\beta$ limit become suppressed. One of the most significant suppressions is of proton decay from dimension 5 operators.

 In the MSSM, dimension 5 operators in the superpotential of the form $qqq\ell$ and $\bar u \bar u \bar d \bar e$ can lead to rapid proton decay and must be quite suppressed. These are dangerous even if associated with an inverse power of the Planck scale and with the same small parameters which suppress the light quark and lepton masses. However in the $U(1)_R$ symmetric limit, such supersymmetric operators do not lead to proton decay, since
a linear combination of baryon number and $U(1)_R$ charge remains unbroken. The proton decay rate is thus suppressed by a welcome factor of the square of ratio of the R symmetry breaking  terms to the R symmetric supersymmetry breaking terms, a factor of approximately $1/\tan^2\beta\sim  4 \times 10^{-4}$.

%----------------------------------------------------------------------------
\section{Unification of couplings}
\label{unify}
One rational for
supersymmetry is coupling constant unification. If we add matter to
the MSSM in incomplete multiplets under the unifying group
the usual successful prediction of $s_W^2\approx .23$ may be lost. 
In the \nomu\ we have added matter in the adjoint representation of
$U(1)\otimes SU(2)\otimes SU(3)$, which will not preserve the usual  prediction. It
is, however a simple matter to embed the $T,S$ and $O$ fields into a complete
adjoint multiplet of a GUT such as $SU(3)^3$\cite{Rizov:1981dp} or $SU(5)$
\cite{Georgi:1974sy}.

 It would be economical, although not necessary, to have the other fields of the
multiplet serve as the messenger fields of a gauge mediated model.Thus we could take the messenger multiplets  to transform under the standard $SU(3)\otimes SU(2)\otimes U(1)$ gauge group as any number of complete unified multiplets plus in addition a  multiplet transforming either as 
\be
(1,2, \pm 1/2)+(1,1,\pm1)+(1,1,\pm1)
\ee
or as \be(3,2,-5/6)+(\bar 3,2,5/6)\  .\ee

There is  an upper
bound on the mass scale  of the new fields for  unification will be
preserved, with the usual one loop result being obtained in the limit where all fields are at the electroweak scale. However the resulting
constraint  is very mild.

If one assumes all the \nomu\ superpartners are at the weak scale,
and computes the one-loop running neglecting threshold effects, one
can  fit the scales of the new matter multiplet and GUT  to the low energy
gauge coupling constants. The result is
\beq
M_{\rm new}&=&M_{\rm weak}e^{{2\pi\over
    3}\left({12\over\alpha_2}-{5\over\alpha_1}-{7\over\alpha_3}\right)}\y \\
M_{\rm GUT}&=&M_{\rm weak}e^{{5\pi\over
    6}\left({3\over\alpha_2}-{1\over\alpha_1}-{2\over\alpha_3}\right)}\ .
\eeq
By taking values for the coupling constants at the edge of their allowed
ranges, {\it e.g.} $\alpha(M_Z)$ = 1/127.7, $\alpha_s(M_Z)$ =
0.122, and $s_W^2$ = 0.233 the additional matter fields can be as heavy as $3\times 10^7$ GeV
and the GUT scale as high as $10^{18}$ GeV. Threshold effects at the GUT, messenger
and \nomu\ scales  and higher loop corrections   make order one
changes in these predictions. This constraint is less stringent than the upper bound on the messenger scale found in section~\ref{rsgmsb}, and so if the additional fields are part of the messenger sector the gauge couplings will unify, to a good approximation.

It is tempting to speculate on an extra dimensional origin for such split adjoint matter multiplets. 
After all, extra dimensional theories in which gauge bosons live in the bulk  and chiral matter fields live on a three brane typically have additional matter fields in the adjoint representation when described four dimensionally, unless the extra dimension is orbifolded. The adjoint fields might be 
$N=2$ superpartners of the gauge fields.
We will leave aside such model building issues here, but these will be explored in ref. \cite{fox}. 

%----------------------------------------------------------------------------
\section{Summary}
\label{conclusions}

We have proposed a viable supersymmetric model without a $\mu$ parameter,
by extending the matter content of the MSSM. Charginos, neutralinos
and gluinos get supersymmetry breaking Dirac mass terms by mixing 
with, respectively, the fermionic components of a $SU(2)$ triplet chiral superfield, a singlet chiral superfield, 
and a color octet. 
The \nomu\ can naturally arise from either gauge or gravity mediation, 
if the supersymmetry breaking sector respects an approximate $U(1)_R$
symmetry. Such an approximate symmetry can easily arise by accident, as a consequence of the absence of  gauge singlet chiral superfields with $F-$terms in the supersymmetry breaking or mediation sector.

We have studied aspects of the phenomenology of the \nomu, which contains two 
light charginos with masses within the reach of the TeVatron, and 
 a quasi-Dirac neutralino lighter than the $Z$ boson. Such a light quasi-Dirac particle might have some unusual features as a dark matter candidate, along the lines of \cite{Smith:2001hy}.

A very strong constraint on the \nomu\ model comes from the contribution
to the electroweak $T$ parameter. In the absence of a $\mu$ term, the 
superpotential coupling $h_T$ in (\ref{tcoupling}) should be large
for the charginos to be heavier than the current experimental limit 
from LEP II. This coupling breaks custodial SU(2) symmetry and 
gives the main contribution to the $T$ parameter. 
The size of this correction is directly correlated with the chargino 
mass bounds, so it would be very interesting to find the experimental bounds on the chargino masses in this 
particular model, 
since the existing ones for the MSSM are not applicable.

We have also computed the supersymmetric contribution to the muon 
anomalous magnetic moment within the \nomu. We find a strong 
suppression due to the approximate $U(1)_R$ symmetry; as a consequence, 
the effect is small even for large values of $\tan\beta$ and light superpartners. 

The MSSM bound on the lightest Higgs mass does not apply, though, since 
the scalar sector is also enlarged by the scalar components of the 
$SU(2)$ triplet and scalar chiral superfields, and there are new, $F-$component contributions to the Higgs quartic coupling. There will still be some upper bounds from triviality of the superpotential couplings, as computed
in  general models with Higgs triplets in refs. \cite{Espinosa:1992wt,Espinosa:1998re}.

There are many aspects of this model deserving further study such as the most effective search strategies for the superpartners, a more complete, predictive  and explicit gauge mediated origin for the supersymmetry breaking parameters, the origin of the approximate $U(1)_R$ symmetry and of the small explicit breaking terms, and whether the required fields and symmetries could be obtained from extra dimensions.

%----------------------------------------------------------------------------

\setcounter{equation}{0}
\renewcommand{\theequation}{A. \arabic{equation}}

\section{Appendix} 

The left- and right- muon-smuon-neutralino and muon-sneutrino-chargino 
vertices which appear in $\delta a_\mu$, eqs. (\ref{deltan}), 
(\ref{deltac}), are given by 
\beq
n_{im}^R&=&\sqrt{2} N_{i3} X_{m2}+y_{\mu} N_{i5} X_{m1}
\\
n_{im}^L&=&\frac{1}{\sqrt{2}} (g_2 N_{i4}+g_1 N_{i3}) X^*_{m1}-y_{\mu} N_{i5}
X^*_{m2}
\\
c_k^R&=&y_{\mu} U_{k3}
\\
c_k^L&=&-g_2 V_{k2}
\eeq
where $N,U,V,X$ are the unitary transformations that diagonalize the 
neutralino, chargino and smuon mass matrices, i.e. they 
satisfy 
\beq
N^* M_{\chi^0} N^{\dagger} &=& 
{\rm diag}(m_{\chi^0_1},m_{\chi^0_2},m_{\chi^0_3},m_{\chi^0_4},m_{\chi^0_5},
m_{\chi^0_6}) 
\\
U^* M_{\chi^{\pm}} V^{\dagger}&=& 
{\rm diag}(m_{\chi^{\pm}_1},m_{\chi^{\pm}_2},m_{\chi^{\pm}_3})
 \\
X M^2_{\tilde{\mu}} X^{\dagger}&=& 
{\rm diag}(m^2_{\tilde{\mu}_1}m^2_{\tilde{\mu}_2})
\eeq

The loop functions have the form
\beq
F_1^N(x)&=&\frac{2}{(1-x)^4} (1-6x+3x^2+2x^3-6x^2 \log x) \\
F_2^N(x)&=&\frac{3}{(1-x)^3}(1-x^2+2x\log x)\\
F_1^C(x)&=&\frac{2}{(1-x)^4}(2+3x-6x^2+x^3+6x\log x)\\
F_2^C(x)&=&\frac{-3}{2(1-x)^3} (3-4x+x^2+2 \log x)      
\eeq
and are normalized so that $F_i^{N}(1) = F_i^{C}(1)$=1 ($i=1,2$), 
corresponding to degenerate sparticles.

\vskip 0.25in
{\bf Acknowledgments} 
\vskip 0.15in
A.N. thanks Neal Weiner for useful conversations. 
N.R. and V.S. thank Oscar Vives for discussions. 
This work was partially 
supported by the DOE under contract DE-FGO3-96-ER40956, by the  
Spanish MCyT grants PB98-0693 and FPA2001-3031, 
by ERDF funds from the European Commission, and by  
the TMR network contract HPRN-CT-2000-00148 of the European Union. 
\bibliography{rsgmsb}

\providecommand{\href}[2]{#2}\begingroup\raggedright\begin{thebibliography}{10}

\bibitem{Nelson:1993vc}
A.~E. Nelson and L.~Randall, {\it Naturally large tan beta},  {\em Phys. Lett.}
  {\bf B316} (1993) 516--520,
  [\href{http://xxx.lanl.gov/abs/http://arXiv.org/abs/hep-ph/9308277}{{\tt
  http://arXiv.org/abs/hep-ph/9308277}}].

\bibitem{Randall:1997zi}
L.~Randall, {\it New mechanisms of gauge-mediated supersymmetry breaking},
  {\em Nucl. Phys.} {\bf B495} (1997) 37--56,
  [\href{http://xxx.lanl.gov/abs/http://arXiv.org/abs/hep-ph/9612426}{{\tt
  http://arXiv.org/abs/hep-ph/9612426}}].

\bibitem{Csaki:1998if}
C.~Csaki, L.~Randall, and W.~Skiba, {\it Composite intermediary and mediator
  models of gauge- mediated supersymmetry breaking},  {\em Phys. Rev.} {\bf
  D57} (1998) 383--390,
  [\href{http://xxx.lanl.gov/abs/http://arXiv.org/abs/hep-ph/9707386}{{\tt
  http://arXiv.org/abs/hep-ph/9707386}}].

\bibitem{Poppitz:1997xw}
E.~Poppitz and S.~P. Trivedi, {\it Some remarks on gauge-mediated supersymmetry
  breaking},  {\em Phys. Lett.} {\bf B401} (1997) 38--46,
  [\href{http://xxx.lanl.gov/abs/http://arXiv.org/abs/hep-ph/9703246}{{\tt
  http://arXiv.org/abs/hep-ph/9703246}}].

\bibitem{Arkani-Hamed:1998kj}
N.~Arkani-Hamed, G.~F. Giudice, M.~A. Luty, and R.~Rattazzi, {\it
  Supersymmetry-breaking loops from analytic continuation into superspace},
  {\em Phys. Rev.} {\bf D58} (1998) 115005,
  [\href{http://xxx.lanl.gov/abs/http://arXiv.org/abs/hep-ph/9803290}{{\tt
  http://arXiv.org/abs/hep-ph/9803290}}].

\bibitem{fox}
P.~Fox, A.~Nelson, and N.~Weiner, {\it Dirac gaugino masses and supersoft
  supersymmetry breaking},
  \href{http://xxx.lanl.gov/abs/http://arXiv.org/abs/hep-ph/0206096}{{\tt
  http://arXiv.org/abs/hep-ph/0206096}}.

\bibitem{Dine:1995vc}
M.~Dine, A.~E. Nelson, and Y.~Shirman, {\it Low-energy dynamical supersymmetry
  breaking simplified},  {\em Phys. Rev.} {\bf D51} (1995) 1362--1370,
  [\href{http://xxx.lanl.gov/abs/http://arXiv.org/abs/hep-ph/9408384}{{\tt
  http://arXiv.org/abs/hep-ph/9408384}}].

\bibitem{Hall:1991hq}
L.~J. Hall and L.~Randall, {\it U(1)-r symmetric supersymmetry},  {\em Nucl.
  Phys.} {\bf B352} (1991) 289--308.

\bibitem{Randall:1992cq}
L.~Randall and N.~Rius, {\it The minimal U(1)-r symmetric model revisited},
  {\em Phys. Lett.} {\bf B286} (1992) 299--306.

\bibitem{Dine:1992yw}
M.~Dine and D.~MacIntire, {\it Supersymmetry, naturalness, and dynamical
  supersymmetry breaking},  {\em Phys. Rev.} {\bf D46} (1992) 2594--2601,
  [\href{http://xxx.lanl.gov/abs/http://arXiv.org/abs/hep-ph/9205227}{{\tt
  http://arXiv.org/abs/hep-ph/9205227}}].

\bibitem{Feng:1996dn}
J.~L. Feng, N.~Polonsky, and S.~Thomas, {\it The light higgsino-gaugino
  window},  {\em Phys. Lett.} {\bf B370} (1996) 95--105,
  [\href{http://xxx.lanl.gov/abs/http://arXiv.org/abs/hep-ph/9511324}{{\tt
  http://arXiv.org/abs/hep-ph/9511324}}].

\bibitem{Dine:1996ag}
M.~Dine, A.~E. Nelson, Y.~Nir, and Y.~Shirman, {\it New tools for low-energy
  dynamical supersymmetry breaking},  {\em Phys. Rev.} {\bf D53} (1996)
  2658--2669,
  [\href{http://xxx.lanl.gov/abs/http://arXiv.org/abs/hep-ph/9507378}{{\tt
  http://arXiv.org/abs/hep-ph/9507378}}].

\bibitem{Poppitz:1996fh}
E.~Poppitz and S.~P. Trivedi, {\it Some examples of chiral moduli spaces and
  dynamical supersymmetry breaking},  {\em Phys. Lett.} {\bf B365} (1996)
  125--131,
  [\href{http://xxx.lanl.gov/abs/http://arXiv.org/abs/hep-th/9507169}{{\tt
  http://arXiv.org/abs/hep-th/9507169}}].

\bibitem{Randall:1998uk}
L.~Randall and R.~Sundrum, {\it Out of this world supersymmetry breaking},
  {\em Nucl. Phys.} {\bf B557} (1999) 79--118,
  [\href{http://xxx.lanl.gov/abs/http://arXiv.org/abs/hep-th/9810155}{{\tt
  http://arXiv.org/abs/hep-th/9810155}}].

\bibitem{Giudice:1998xp}
G.~F. Giudice, M.~A. Luty, H.~Murayama, and R.~Rattazzi, {\it Gaugino mass
  without singlets},  {\em JHEP} {\bf 12} (1998) 027,
  [\href{http://xxx.lanl.gov/abs/http://arXiv.org/abs/hep-ph/9810442}{{\tt
  http://arXiv.org/abs/hep-ph/9810442}}].

\bibitem{Groom:2000in}
{\bf Particle Data Group} Collaboration, D.~E. Groom {\em et.~al.}, {\it Review
  of particle physics},  {\em Eur. Phys. J.} {\bf C15} (2000) 1--878.

\bibitem{Chivukula:2000fe}
R.~S. Chivukula, {\it Limits on the mass of a composite higgs boson},
  \href{http://xxx.lanl.gov/abs/http://arXiv.org/abs/hep-ph/0005168}{{\tt
  http://arXiv.org/abs/hep-ph/0005168}}.

\bibitem{Brown:2001mg}
{\bf Muon g-2} Collaboration, H.~N. Brown {\em et.~al.}, {\it Precise
  measurement of the positive muon anomalous magnetic moment},  {\em Phys. Rev.
  Lett.} {\bf 86} (2001) 2227--2231,
  [\href{http://xxx.lanl.gov/abs/http://arXiv.org/abs/hep-ex/0102017}{{\tt
  http://arXiv.org/abs/hep-ex/0102017}}].

\bibitem{Davier:1998si}
M.~Davier and A.~Hocker, {\it New results on the hadronic contributions to
  alpha(m(z)**2) and to (g-2)(mu)},  {\em Phys. Lett.} {\bf B435} (1998)
  427--440,
  [\href{http://xxx.lanl.gov/abs/http://arXiv.org/abs/hep-ph/9805470}{{\tt
  http://arXiv.org/abs/hep-ph/9805470}}].

\bibitem{Jegerlehner:2001ca}
F.~Jegerlehner, {\it The effective fine structure constant at Tesla energies},
  \href{http://xxx.lanl.gov/abs/http://arXiv.org/abs/hep-ph/0105283}{{\tt
  http://arXiv.org/abs/hep-ph/0105283}}.

\bibitem{Hayakawa:2001bb}
M.~Hayakawa and T.~Kinoshita, {\it Comment on the sign of the pseudoscalar pole
  contribution to the muon g-2},
  \href{http://xxx.lanl.gov/abs/http://arXiv.org/abs/hep-ph/0112102}{{\tt
  http://arXiv.org/abs/hep-ph/0112102}}.

\bibitem{Knecht:2001qg}
M.~Knecht, A.~Nyffeler, M.~Perrottet, and E.~De~Rafael, {\it Hadronic
  light-by-light scattering contribution to the muon g-2: An effective field
  theory approach},  {\em Phys. Rev. Lett.} {\bf 88} (2002) 071802,
  [\href{http://xxx.lanl.gov/abs/http://arXiv.org/abs/hep-ph/0111059}{{\tt
  http://arXiv.org/abs/hep-ph/0111059}}].

\bibitem{Knecht:2001qf}
M.~Knecht and A.~Nyffeler, {\it Hadronic light-by-light corrections to the muon
  g-2: The pion-pole contribution},  {\em Phys. Rev.} {\bf D65} (2002) 073034,
  [\href{http://xxx.lanl.gov/abs/http://arXiv.org/abs/hep-ph/0111058}{{\tt
  http://arXiv.org/abs/hep-ph/0111058}}].

\bibitem{Marciano:2001qq}
W.~J. Marciano and B.~L. Roberts, {\it Status of the hadronic contribution to
  the muon (g-2) value},
  \href{http://xxx.lanl.gov/abs/http://arXiv.org/abs/hep-ph/0105056}{{\tt
  http://arXiv.org/abs/hep-ph/0105056}}.

\bibitem{Bartos:2001pg}
E.~Bartos, A.~Z. Dubnickova, S.~Dubnicka, E.~A. Kuraev, and E.~Zemlyanaya, {\it
  Scalar and pseudoscalar meson pole terms in the hadronic light-by-light
  contributions to muon a(mu)(had)},
  \href{http://xxx.lanl.gov/abs/http://arXiv.org/abs/hep-ph/0106084}{{\tt
  http://arXiv.org/abs/hep-ph/0106084}}.

\bibitem{Bijnens:2001cq}
J.~Bijnens, E.~Pallante, and J.~Prades, {\it Comment on the pion pole part of
  the light-by-light contribution to the muon g-2},  {\em Nucl. Phys.} {\bf
  B626} (2002) 410--411,
  [\href{http://xxx.lanl.gov/abs/http://arXiv.org/abs/hep-ph/0112255}{{\tt
  http://arXiv.org/abs/hep-ph/0112255}}].

\bibitem{Blokland:2001pb}
I.~Blokland, A.~Czarnecki, and K.~Melnikov, {\it Pion pole contribution to
  hadronic light-by-light scattering and muon anomalous magnetic moment},  {\em
  Phys. Rev. Lett.} {\bf 88} (2002) 071803,
  [\href{http://xxx.lanl.gov/abs/http://arXiv.org/abs/hep-ph/0112117}{{\tt
  http://arXiv.org/abs/hep-ph/0112117}}].

\bibitem{Melnikov:2001uw}
K.~Melnikov, {\it On the theoretical uncertainties in the muon anomalous
  magnetic moment},  {\em Int. J. Mod. Phys.} {\bf A16} (2001) 4591--4612,
  [\href{http://xxx.lanl.gov/abs/http://arXiv.org/abs/hep-ph/0105267}{{\tt
  http://arXiv.org/abs/hep-ph/0105267}}].

\bibitem{Ramsey-Musolf:2002cy}
M.~Ramsey-Musolf and M.~B. Wise, {\it Hadronic light-by-light contribution to
  muon g-2 in chiral perturbation theory},
  \href{http://xxx.lanl.gov/abs/http://arXiv.org/abs/hep-ph/0201297}{{\tt
  http://arXiv.org/abs/hep-ph/0201297}}.

\bibitem{Moroi:1996yh}
T.~Moroi, {\it The muon anomalous magnetic dipole moment in the minimal
  supersymmetric standard model},  {\em Phys. Rev.} {\bf D53} (1996)
  6565--6575,
  [\href{http://xxx.lanl.gov/abs/http://arXiv.org/abs/hep-ph/9512396}{{\tt
  http://arXiv.org/abs/hep-ph/9512396}}].

\bibitem{Martin:2001st}
S.~P. Martin and J.~D. Wells, {\it Muon anomalous magnetic dipole moment in
  supersymmetric theories},  {\em Phys. Rev.} {\bf D64} (2001) 035003,
  [\href{http://xxx.lanl.gov/abs/http://arXiv.org/abs/hep-ph/0103067}{{\tt
  http://arXiv.org/abs/hep-ph/0103067}}].

\bibitem{Rizov:1981dp}
V.~A. Rizov, {\it A gauge model of the electroweak and strong interactions
  based on the group su(3)-l x su(3)-r x su(3)-c},  {\em Bulg. J. Phys.} {\bf
  8} (1981) 461--477.

\bibitem{Georgi:1974sy}
H.~Georgi and S.~L. Glashow, {\it Unity of all elementary particle forces},
  {\em Phys. Rev. Lett.} {\bf 32} (1974) 438--441.

\bibitem{Smith:2001hy}
D.~R. Smith and N.~Weiner, {\it Inelastic dark matter},  {\em Phys. Rev.} {\bf
  D64} (2001) 043502,
  [\href{http://xxx.lanl.gov/abs/http://arXiv.org/abs/hep-ph/0101138}{{\tt
  http://arXiv.org/abs/hep-ph/0101138}}].

\bibitem{Espinosa:1992wt}
J.~R. Espinosa and M.~Quiros, {\it Higgs triplets in the supersymmetric
  standard model},  {\em Nucl. Phys.} {\bf B384} (1992) 113--146.

\bibitem{Espinosa:1998re}
J.~R. Espinosa and M.~Quiros, {\it Gauge unification and the supersymmetric
  light higgs mass},  {\em Phys. Rev. Lett.} {\bf 81} (1998) 516--519,
  [\href{http://xxx.lanl.gov/abs/http://arXiv.org/abs/hep-ph/9804235}{{\tt
  http://arXiv.org/abs/hep-ph/9804235}}].

\end{thebibliography}\endgroup
\bibliographystyle{JHEP}
\end{document}